\documentclass[10pt,conference]{IEEEtran}

\usepackage{epsfig, setspace, amsmath, epsf, amssymb, bm, theorem, cite, graphicx, epstopdf, algorithm, algpseudocode, float, color, accents}
\usepackage{multirow}
\usepackage{authblk}
\usepackage[table,xcdraw]{xcolor}
\usepackage{mathtools}

\newtheorem{theorem}{Theorem}

\newtheorem{remark}{Remark}
\newtheorem{lemma}{Lemma}

\newenvironment{Proof}[1]{\medskip\par\noindent{\bf Proof:\,}\,#1}{{\mbox{\,$\blacksquare$}\par}}

\IEEEoverridecommandlockouts
\newcommand{\bbE}{{\mathbb{E}}}

\allowdisplaybreaks

\begin{document}


\title{Timely Multi-Process Estimation with Erasures\thanks{This work was supported by the U.S. National Science Foundation under Grants CNS 21-14537 and ECCS 21-46099.}}

\author[1]{Karim Banawan}
\author[2]{Ahmed Arafa}
\author[3]{Karim G. Seddik}
\affil[1]{\normalsize Department of Electrical Engineering, Alexandria University, Egypt.}
\affil[2]{\normalsize Department of Electrical and Computer Engineering, University of North Carolina at Charlotte, USA}
\affil[3]{\normalsize Electronics and Communications Engineering Department, American University in Cairo, Egypt.}

\maketitle

\begin{abstract}
We consider a multi-process remote estimation system observing $K$ independent Ornstein-Uhlenbeck processes. In this system, a shared sensor samples the $K$ processes in such a way that the long-term average sum mean square error (MSE) is minimized. The sensor operates under a total sampling frequency constraint $f_{\max}$ and samples the processes according to a Maximum-Age-First (MAF) schedule. The samples from all processes consume random processing delays, and then are transmitted over an erasure channel with probability $\epsilon$. Aided by optimal structural results, we show that the optimal sampling policy, under some conditions, is a \emph{threshold policy}. We characterize the optimal threshold and the corresponding optimal long-term average sum MSE as a function of $K$, $f_{\max}$, $\epsilon$, and the statistical properties of the observed processes. 
\end{abstract}

\section{Introduction}

We study the problem of timely tracking of multiple random processes using shared resources. This setting arises in many practical situations of remote estimation applications. Recent works have drawn connections between the quality of the estimates at the destination, measured through mean square error (MSE), and the {\it age of information} (AoI) metric that assesses timeliness and freshness of the received data, see, e.g., the survey in \cite[Section~VI]{aoi-survey-jsac}. We extend these results to {\it multi-process} estimation settings in this work.

AoI is defined as the time elapsed since the latest received message has been generated at its source. It has been studied extensively in the past few years in various contexts, see, e.g., \cite{yates_age_1, ephremides_age_random, ephremides_age_management, ephremides_age_non_linear, yates-age-mltpl-src, talak-aoi-delay, inoue-aoi-general-formula-fcfs, soysal-aoi-gg11, zou-waiting-aoi, modiano-age-bc, sun-age-mdp, zhou-age-iot, sun-cyr-aoi-non-linear, tang-aoi-power-multi-state, yates_age_eh, jing-age-online, baknina-updt-info, arafa-age-online-finite, bacinoglu-aoi-eh-finite-gnrl-pnlty, leng-aoi-eh-cog-radio, batu-aoi-multihop, bedewy-aoi-multihop, himanshu-age-source-coding, zhang-arafa-aoi-pricing-wiopt, arafa-aoi-compute, yang-arafa-aoi-fl}. Relevant to this work is the fact that AoI can be closely tied to MSE in random processes tracking applications. The works in \cite{klugel2019aoi-fr, mitra-estimation-graphs-aoi, chakravorty-estimation-pckt-drop-markov} characterize implicit and explicit relationships between MSE and AoI under different estimation contexts. References \cite{ayan-aoi-voi-cntrl, roth-mse-aoi-finite-blocklength}, however, consider the notion of the value of information (mainly through MSE) and show that optimizing it can be different from optimizing AoI. Lossy source coding and distorted updates for AoI minimization is considered in \cite{ramirez-aoi-compression, bastopcu-aoi-distortion, bastopcu-partial-updates}. The notion of age of incorrect information (AoII) is introduced in in \cite{maatouk-aoii}, adding more context to AoI by capturing erroneous updates. The works in \cite{sun-wiener, OU_original} consider sampling of Wiener and Ornstein-Uhlenbeck (OU) processes for the purpose of remote estimation, and draw connections between MSE and AoI. Our recent work in \cite{sample_quantize} also focuses on characterizing the relationship of MSE and AoI, yet with the additional presence of coding and quantization. Reference \cite{guo2021optimal} shows the optimality of threshold policies for tracking OU processes under rate constraints.

Reference \cite{OU_original} is closely related to our setting, in which optimal sampling methods to minimize the long-term average MSE for an OU process is derived. It is shown that if sampling times are independent of the instantaneous values of the process (signal-independent sampling) the minimum MSE (MMSE) reduces to an increasing function of AoI (age penalty). Then, threshold policies are shown optimal in this case, in which a new sample is acquired only if the expected age-penalty surpasses a certain value. This paper extends \cite{OU_original} (and the related studies in \cite{sample_quantize,guo2021optimal}) to multiple OU processes.

\begin{figure}[t]
\center
\includegraphics[scale=.4]{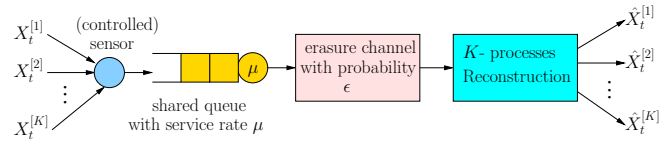}
\vspace{-.1in}
\caption{System model.}
\label{fig:system model}
\vspace{-0.22in}
\end{figure}

In this paper, we study a remote sensing problem consisting of a shared controlled sensor, a shared queue, and a receiver (see Fig.~\ref{fig:system model}) to track $K$ independent, but not necessarily identical, OU processes.\footnote{The OU process is the continuous-time analogue of the first-order autoregressive process \cite{ou-brownian-motion, doob-brownian-motion}, and is used to model various physical phenomena, and has relevant applications in control and finance.} The sensor transmits the collected samples over an erasure channel with probability $\epsilon$ after being processed for a random delay with service rate $\mu$. The sensor generates the samples \emph{at will}, subject to a total sampling frequency constraint $f_{\max}$. The goal is to minimize the long-term average sum MSE of the $K$ processes. We focus on \emph{maximum-age-first} (MAF) scheduling, where the scheduler chooses the process with the largest AoI to be sampled. MAF scheduling results in obtaining a \emph{fresh} sample from the \emph{same} process until an unerased sample from that process is conveyed to the receiver. We show that the optimal stationary deterministic policy is a \emph{threshold policy}. We characterize the optimal threshold $\tau^*(K,f_{\max},\epsilon,\boldsymbol{\theta},\boldsymbol{\sigma})$ and the corresponding long-term average sum MSE in terms of the processes statistical properties ($\boldsymbol{\theta},\boldsymbol{\sigma}$), $\epsilon$, and $f_{\max}$. The threshold is a maximum of two threshold values: one due to a nonbinding sampling frequency constraint scenario, and another due to a binding scenario. Our numerical results show that 1) the optimal threshold $\tau^*$ is an increasing function in the erasure probability $\epsilon$, and 2) the optimal threshold is an increasing function in the number of the observed processes $K$.

\section{System Model}

We consider a sensing system in which $K$ independent, but not necessarily identical, OU processes are remotely monitored using a {\it shared} sensor that transmits samples from the processes over an erasure channel to a receiver. Denote the $k$th process value at time $s$ by $X_s^{[k]}$. Given $X_s^{[k]}$, the $k$th OU process evolves, for $t\geq s$, as \cite{ou-brownian-motion, doob-brownian-motion}
\begin{align} \label{eq_ou_evol}
X_t^{[k]}=X_s^{[k]}e^{-\theta_k(t-s)}+\frac{\sigma_k}{\sqrt{2\theta_k}}e^{-\theta_k(t-s)}W_{e^{2\theta_k(t-s)}-1},
\end{align} 
where $W_t$ denotes a Wiener process, while $\theta_k>0$ and $\sigma_k>0$ are fixed parameters that control how fast the process evolves. We assume that the processes are initiated as $X_0^{[k]}\sim\mathcal{N}\left(0,\sigma_k^2/2\theta_k\right)$.\footnote{This way, the variance of $X_t^{[k]}$ is $\sigma_k^2/2\theta_k,~\forall t$.}

To estimate $\left\{X_t^{[k]}\right\}$ at the receiver, the sensor observes the $k$th OU process at specific time instants $\left\{S_i^{[k]}\right\}$ and sends the samples to the receiver. Sampling instants are fully-controlled, i.e., samples are {\it generated-at-will}. We focus on \emph{signal-independent} sampling policies, in which the optimal sampling instants depend on the statistical measures of the processes and not on exact processes' values.

The sensor must obey a {\it total sampling frequency constraint} $f_{\max}$. Let $\ell_i$ denote the $i$th sampling instance {\it regardless} of the identity of the process being sampled. Hence, the sampling constraint is expressed as follows:
\begin{align} \label{eq_smpl-const-genatwill}
    \liminf_{n\rightarrow\infty}\frac{1}{n}\mathbb{E}\left[\sum_{i=1}^n\ell_{i+1}-\ell_i\right]\geq\frac{1}{f_{\max}},
\end{align}
which indicates that the sensor shares the sampling budget $f_{\max}$ among the $K$ processes. Samples go through a shared processing queue, whose service model follows a Poisson process with service rate $\mu$, i.e., service times are independent and identically distributed (i.i.d.)~$\sim\exp(\mu)$ across samples. Served samples, however, are prune to erasures with probability $\epsilon$, also independently across samples. Immediate erasure status feedback is available.

Samples are time-stamped prior to transmissions, and successfully received samples from process $k$ determine the {\it age-of-information} (AoI) of that process at the receiver, denoted $\texttt{AoI}^{[k]}(t)$. AoI is defined as the time elapsed since the latest successfully received sample's time stamp. We focus on \emph{Maximum-Age-First (MAF) scheduling}, in which the processes are sampled according to their relative AoI's, with priority given to the process with highest AoI. Hence, at time $t$, process
	\begin{align}
	    \kappa(t)\triangleq\arg\max_k\texttt{AoI}^{[k]}(t)    
	\end{align}
is sampled. Observe that the value of $\kappa(t)$ will {\it not} change unless a successful transmission occurs. Therefore, in case of erasure events, a {\it fresh} sample is generated from the {\it same} process being served and transmission is re-attempted. Under MAF scheduling, and since the channel behaves similarly for all processes, each process will be given an equal share of the allowed sampling budget, i.e., each process will be sampled at a rate of $f_{\max}/K$, and the sampling constraint in (\ref{eq_smpl-const-genatwill}) becomes
	\begin{align} \label{eq_smpl-const-maf}
        \liminf_{n\rightarrow\infty}\frac{1}{n}\mathbb{E}\left[\sum_{i=1}^nS_{i+1}^{[k]}-S_i^{[k]}\right]\geq\frac{K}{f_{\max}},\quad\forall k.
    \end{align}
    
Let $\tilde{S}_i^{[k]}$ denote the sampling instance of the $i$th {\it successfully} received sample from the $k$th process, and (re-)define $S_i^{[k]}(m)$ as the sampling instance of the $m$th {\it attempt} to convey the $i$th sample of the $k$th process, $m=1,\dots,M_i^{[k]}$, with $M_i^{[k]}$ denoting the number of trials. Hence, we have $S_i^{[k]}(m)\leq\tilde{S}_i^{[k]},~\forall m$, with equality at $m=M_i^{[k]}$. Our channel model indicates that $M_i^{[k]}$'s are i.i.d.~$\sim \text{geometric}(1-\epsilon)$. Each sample $X_{S_i^{[k]}(m)}^{[k]}$ incurs a service time of $Y_i^{[k]}(m)$ time units with $Y_i^{[k]}(m)$'s being i.i.d.~$\sim\exp(\mu)$. The successfully received sample, $X_{\tilde{S}_i^{[k]}}^{[k]}$, arrives at the receiver at time $D_i^{[k]}$, i.e.,
\begin{align}
D_i^{[k]}=\tilde{S}_i^{[k]}+Y_i^{[k]}\left(M_i^{[k]}\right).
\end{align}
Based on this notation, one can characterize the AoI of the $k$th OU process as follows:
\begin{align} \label{eq_aoi}
\texttt{AoI}^{[k]}(t)=t-\tilde{S}_i^{[k]}, \qquad D_{i}^{[k]} \leq t < D_{i+1}^{[k]}. 
\end{align}

The receiver collects the unerased samples from all processes and uses them to construct minimum mean square error (MMSE) estimates. Since the processes are independent, and by the strong Markov property of the OU process, the MMSE estimate for the $k$th process at time $t$, denoted $\hat{X}_t^{[k]}$, is based solely on the latest successfully received sample from that process. Thus, for $D_{i}^{[k]} \leq t < D_{i+1}^{[k]}$, we have \cite{OU_original,sample_quantize}
\begin{align}
\hat{X}_t^{[k]}=\mathbb{E}\left[X_t^{[k]}\Big|\tilde{S}_i^{[k]}, X_{\tilde{S}_i^{[k]}}\right] \stackrel{(\ref{eq_ou_evol})}{=}X_{\tilde{S}_i^{[k]}}e^{-\theta_k(t-\tilde{S}_i^{[k]})}.
\end{align}
Hence, the instantaneous mean square error (MSE) in estimating the $k$th process at time $t\in\left[D_i^{[k]},D_{i+1}^{[k]}\right)$ is \cite{OU_original,sample_quantize}
\begin{align}
\texttt{mse}^{[k]}\left(t,\tilde{S}_i^{[k]}\right)&\triangleq\bbE\left[\left(X_t^{[k]}-\hat{X}_t^{[k]}\right)^2\right] \\ 
&=\frac{\sigma_k^2}{2\theta_k}\left(1-e^{-2\theta_k\left(t-\tilde{S}_i^{[k]}\right)}\right), \label{MSE_value}
\end{align}
which is an increasing function of the AoI in (\ref{eq_aoi}). 
Next, we define the long-term time average MSE of the $k$th process as
\begin{align}\label{long_term_MSE}
\overline{\texttt{mse}^{[k]}}\!\triangleq\!\limsup_{T \rightarrow \infty}\frac{\sum_{i=1}^T \bbE\left[\int_{D_i^{[k]}}^{D_{i+1}^{[k]}} \texttt{mse}^{[k]}\left(t,\tilde{S}_i^{[k]}\right)dt\right]}{\sum_{i=1}^T\mathbb{E}\left[ D_{i+1}^{[k]}-D_i^{[k]}\right]}.
\end{align}  

Our goal is to choose the sampling instants to minimize a penalty function $g(\cdot)$ of $\left\{\overline{\texttt{mse}^{[k]}}\right\}$. More specifically, to solve
\begin{align} \label{opt_gen_maf}
\min_{\{S_i^{[k]}(m)\}} &\quad 	g\left(\overline{\texttt{mse}^{[1]}}, \cdots, \overline{\texttt{mse}^{[K]}}\right) \nonumber \\
\text{s.t.~~~}  &~~ \liminf_{n\rightarrow\infty}\frac{1}{n}\mathbb{E}\left[\sum_{i=1}^nS_{i+1}^{[k]}-S_i^{[k]}\right]\geq\frac{K}{f_{\max}},~\forall k.
\end{align}

\section{Stationary Policies: Problem Re-Formulation}

In this section, we re-formulate problem \eqref{opt_gen_maf} in terms of a {\it waiting policy} for each process. To that end, we define $W_i^{[k]}(m)$ as the $m$th waiting time before taking the $m$th sample towards conveying the $i$th sample from the $k$th process, $1\leq m\leq M_i^{[k]}$. Without loss of generality, let the MAF schedule be in the order $1,2,\dots,K$. Thus, we have
\begin{align}
    S_i^{[k]}(m)=D_{i}^{[k-1]}&+\sum_{j=1}^{m-1} Y_i^{[k]}(j)+\sum_{j=1}^{m}W_i^{[k]}(j),
\end{align}
with $D_i^{[0]}\triangleq D_{i-1}^{[K]}$. Problem \eqref{opt_gen_maf} now reduces to optimizing the waiting times $\left\{W_i^{[k]}(m)\right\}$. We now define the $i$th \emph{epoch} of the $k$th process, denoted $\Gamma_i^{[k]}$, as the inter-reception time in between its $i$th and $(i+1)$th {\it unerased} samples, i.e.,
\begin{align}
    \Gamma_i^{[k]}=D_{i+1}^{[k]}-D_i^{[k]}.
\end{align}

In this work, we focus on {\it stationary waiting policies} in which the waiting policy $\left\{W_i^{[k]}(m)\right\}$ has the same distribution across all processes' epochs. Note that under MAF scheduling, each process epoch entails a successful transmission of every other process. This, together with the fact that service times and erasure events are i.i.d., induces a stationary distribution across all processes' epochs. Therefore, dropping the indices $i$ and $k$, we have $\Gamma_i^{[k]}\sim\Gamma,~\forall i,k$, where
\begin{align}
\Gamma=\sum_{k=1}^K \sum_{m=1}^{M^{[k]}} W^{[k]}(m) + Y^{[k]}(m).
\end{align}
Now consider a typical epoch for the $k$th process. By stationarity, one can write \eqref{long_term_MSE} as
\begin{align}\label{mse_Gamma}
    \overline{\texttt{mse}^{[k]}}=\frac{ \bbE\left[\int_{D^{[k]}}^{D^{[k]}+\Gamma} \texttt{mse}^{[k]}\left(t,\tilde{S}^{[k]}\right)dt\right]}{\mathbb{E}\left[\Gamma \right]}.
\end{align}
where $D_i^{[k]}\sim D^{[k]}$ and $\tilde{S}_i^{[k]}\sim \tilde{S}^{[k]}$, $\forall i$. In the sequel, {\it we treat the $K$th (last) process's epoch as the typical epoch.} 

In the next lemma, we prove an important structural result, which asserts that the positions of the waiting times do not matter. Specifically, we show that one can achieve the same long-term average MSE penalty by {\it grouping all the waiting times at the beginning of the (typical) epoch.}

\begin{lemma}
Under signal-independent sampling with MAF scheduling and stationary waiting policies, problem \eqref{opt_gen_maf} is equivalent to the following optimization problem:
\begin{align} \label{opt_gen_maf_onewait}
\min_{W \geq 0} &\qquad 	g\left(\overline{\texttt{mse}^{[1]}}, \cdots, \overline{\texttt{mse}^{[K]}}\right) \nonumber \\
\mbox{s.t.} &\qquad \bbE\left[(1-\epsilon)W+\sum_{k=1}^{K}Y^{[k]}\right] \geq \frac{K}{f_{\max}},
\end{align}
where $W\triangleq\sum_{k=1}^K \sum_{m=1}^{M^{[k]}} W^{[k]}(m)$ and the waiting is only performed at the beginning of the epoch.
\end{lemma}

\begin{Proof}
By inspection of the average MSE function in \eqref{mse_Gamma}, since $\Gamma=\sum_{k=1}^K \sum_{m=1}^{M^{[k]}} W^{[k]}(m)+\sum_{k=1}^K \sum_{m=1}^{M^{[k]}} Y^{[k]}(m)$, the waiting times appear in the numerator and denominator as the sum $\sum_{k=1}^K \sum_{m=1}^{M^{[k]}} W^{[k]}(m)$. Thus, for the optimal waiting times $\left\{W^{[k]^*}(m)\right\}$ that solve the optimization problem in \eqref{opt_gen_maf}, the waiting time $W^*=\sum_{k=1}^K\sum_{m=1}^{M^{[k]}} W^{[k]^*}(m)$ achieves the same $\overline{\texttt{mse}^{[k]}}$. Conversely, starting with $W^*$ in the objective function of (\ref{opt_gen_maf_onewait}) and breaking it arbitrarily to any waiting times such that $W^*=\sum_{k=1}^K \sum_{m=1}^{M^{[k]}} W^{[k]^*}(m)$ gives the same objective function in \eqref{opt_gen_maf}.

For the sampling constraint, by observing the telescoping sum in \eqref{eq_smpl-const-maf}, we have that for process $k$,
\begin{align}\label{sampling_constraint}
    \liminf_{n\rightarrow \infty} \frac{1}{n} \bbE\left[\sum_{i=1}^n S_{i+1}^{[k]}-S_i^{[k]}\right]
    &=\liminf_{n\rightarrow \infty} \frac{1}{n} \bbE\left[S_{n+1}^{[k]}\right].
\end{align}
Define $e(n)$ to be the index of the epoch corresponding to the $n$th sample. Hence, we can write the sampling constraint as
\begin{align}
    &\liminf_{n \rightarrow \infty} \frac{e(n)}{n}\cdot \frac{\bbE[S_{n+1}^{[k]}]}{e(n)}\notag\\
    =& \frac{1}{\bbE[M^{[k]}]} \cdot \liminf_{n \rightarrow \infty} \frac{1}{e(n)} \Bigg(\sum_{i=1}^{e(n)-1}\bbE\Bigg[\sum_{k=1}^K  \sum_{m=1}^{M_i^{[k]}}W_i^{[k]}(m)\notag\\
    &\qquad\qquad\qquad\qquad\qquad+Y_i^{[k]}(m)\Bigg]+o\left(e(n)\right)\Bigg)\label{SLLN_M}\\
    =&\frac{1}{\bbE[M^{[k]}]} \cdot\liminf_{n \rightarrow \infty} \frac{1}{e(n)}\!\!  \sum_{i=1}^{e(n)-1}\Bigg( \bbE[W]\notag\\
    &\qquad\qquad\qquad\qquad\quad\left.+ \bbE\left[M_i^{[k]}\right]\cdot \bbE\left[\sum_{k=1}^KY_i^{[k]}\right]\right) \label{wald_sampling}\\
    =&\bbE\left[(1-\epsilon)W+\sum_{k=1}^K Y^{[k]}\right],
\end{align}
where \eqref{SLLN_M} follows from the strong law of large numbers and the fact that the time spent in the $e(n)$th epoch, $\Delta=\sum_{\tilde{k}=1}^{k-1}\sum_{m=1}^{M^{[\tilde{k}]}} W_i^{[\tilde{k}]}(m)+Y_i^{[\tilde{k}]}(m)+\sum_{m=1}^{\tilde{m}} W_i^{[k]}(m)+Y_i^{[k]}(m)$, is $o(e(n))$ and hence $\liminf_{n \rightarrow \infty} \frac{\Delta}{e(n)} =0$, and \eqref{wald_sampling} follows from Wald's identity. 
\end{Proof}

\begin{remark} \label{rmrk_smpl}
Observe that the sampling constraint in problem (\ref{opt_gen_maf_onewait}) will not be active if $f_{\max}>\mu$. This is intuitive since the inter-sampling time, on average, would be larger than the minimum allowable sampling time, controlled by the maximum allowable sampling frequency, in this case. 

If the sampling constraint is binding, which occurs only if $f_{\max}<\mu$, the average waiting time would monotonically increase with the erasure probability $\epsilon$. This is true because no waiting is allowed in between unsuccessful transmissions, whose rate increases with $\epsilon$. Hence, to account for the expected large number of back-to-back sample transmissions in the epoch, one has to wait for a relatively larger amount of time at its beginning so that the sampling constraint is satisfied.
\end{remark}

\section{Optimal Threshold Waiting and Minimum Sum MSE Characterization}

In this section, we provide the optimal solution of problem (\ref{opt_gen_maf_onewait}) for a {\it sum MSE} penalty
\begin{align}
	g\left(\overline{\texttt{mse}^{[1]}}, \cdots, \overline{\texttt{mse}^{[K]}}\right)=\sum_{k=1}^K \overline{\texttt{mse}^{[k]}},
\end{align}
together with a {\it stationary deterministic} waiting policy, in which the waiting value at the beginning of an epoch is given by a deterministic function $w(\cdot)$ of the previous epoch's total service time, denoted $\tilde{Y}$. Note that $\tilde{Y}\sim\sum_{k=1}^{K}\sum_{m=1}^{M^{[k]}} Y^{[k]}(m)$. Such choice of waiting policies emerges naturally since the MSE is an increasing function of the AoI, whose value at the start of the epoch is, in turn, an increasing function of $\tilde{Y}$. Stationary deterministic policies have been used extensively in similar contexts in the literature \cite{sun-wiener, OU_original, sample_quantize} and shown to perform optimally. 

Formally, substituting the above into problem (\ref{opt_gen_maf_onewait}), we now aim at solving the following functional optimization problem:
\begin{align}\label{sumMSE}
    \min_{w(\cdot)\geq 0} &\quad \frac{ \sum_{k=1}^K \bbE\left[\int_{D^{[k]}}^{D^{[k]}+\Gamma} \texttt{mse}^{[k]}\left(t,\tilde{S}^{[k]}\right)dt\right]}{\mathbb{E}\left[\Gamma \right]} \notag\\
    \text{s.t.~} &\quad \bbE\left[w\left(\tilde{Y}\right)\right] \geq \frac{1}{1-\epsilon}\left(\frac{K}{f_{\max}}-\frac{K}{\mu}\right).
\end{align}
Theorem~\ref{thm1} provides the solution of problem \eqref{sumMSE}. A proof sketch is given due to space limits. We use the compact vector notation $\boldsymbol{\theta}\triangleq[\theta_1 \:\: \theta_2 \: \cdots \: \theta_K]$ and $\boldsymbol{\sigma}\triangleq[\sigma_1^2 \:\: \sigma_2^2 \: \cdots \: \sigma_K^2]$.

\begin{theorem}\label{thm1}
The optimal waiting policy $w^*(\cdot)$ that solves problem (\ref{sumMSE}) is given by the \emph{threshold} policy
\begin{align}\label{waiting-policy-opt}
    w^*(z)=\left[\tau^*(K,f_{\max},\epsilon,\boldsymbol{\theta},\boldsymbol{\sigma})-z\right]^+,
\end{align}
where the optimal threshold $\tau^*(K,f_{\max},\epsilon,\boldsymbol{\theta},\boldsymbol{\sigma})$ is given by
\begin{align}\label{threshold}
  \tau^*\!=\!\max\!\left\{\!G_{\boldsymbol{\theta},\boldsymbol{\sigma}}^{-1}\!\left(\beta^*\right)\!, H^{-1}\!\left(\frac{1}{(1\!-\!\epsilon)}\left[ \frac{K}{f_{\max}}\!-\!\frac{K}{\mu}\right]^+\!\right)\!\right\},
\end{align}
where $G_{\boldsymbol{\theta},\boldsymbol{\sigma}}(\tau)\triangleq\sum_{k=1}^K \frac{\sigma_k^2}{2\theta_k}\left(1-\bbE\left[e^{-2\theta_k Y}\right]e^{-2\theta_k\tau}\right)$, and $\beta^*$ corresponds to the optimal long-term average sum MSE in this case, and is given by the unique solution of
\begin{align}
    &\sum_{k=1}^K \frac{\sigma_k^2}{2\theta_k} \left(H(\tau^*)\!+\!\frac{K}{\mu(1\!-\!\epsilon)}\!-\!\frac{1}{2\theta_k}\cdot\frac{\mu}{2\theta_k\!+\!\mu}(1\!-\!F_k(\tau^*))\right)\notag\\
    &\qquad\qquad\qquad\qquad \!-\!\beta^*\left(H(\tau^*)\!+\!\frac{K}{\mu(1\!-\!\epsilon)}\right)=0,
\end{align}
with $H(\cdot)$, and $F_k(\cdot)$ defined in \eqref{H_fn}, and \eqref{F_fn}, respectively.
\begin{figure*}[!t]
	\normalsize
	\setcounter{equation}{29}
	
\begin{align}\label{H_fn}
    H(\tau)=\sum_{\rho=K}^{\infty} \binom{\rho-1}{K-1} \epsilon^{\rho-K} (1-\epsilon)^K \left[\tau \gamma(\mu\tau,\rho) -\frac{\rho}{\mu}\gamma(\mu\tau,\rho+1)\right],
\end{align}
\begin{align}\label{F_fn}
    F_k(\tau)=\sum_{\rho=K}^\infty \binom{\rho-1}{K-1} \epsilon^{\rho-K} (1-\epsilon)^K \left[e^{-2\theta_k \tau} \gamma(\mu \tau, \rho)\!+\!\left(\frac{\mu}{2\theta_k+\mu}\right)^{\rho} (1\!-\!\gamma((2\theta_k\!+\!\mu)\tau, \rho)\right],
\end{align}
where $\gamma(x,y)$ is the normalized incomplete Gamma function defined as $\gamma(x,y)=\frac{1}{(y-1)!} \int_0^x t^{y-1} e^{-t} dt$. 

	\hrulefill
\vspace{-.2in}
\end{figure*}
\end{theorem}

\begin{Proof}[{\it Sketch.}]
We first apply Dinkelbach's approach \cite{dinkelbach-fractional-prog} to transform the fractional objective function of problem \eqref{sumMSE} into a parameterized difference between the numerator and the denominator. This produces the auxiliary optimization problem
\begin{align} \label{opt_aux}
    p(\beta)\triangleq\min_{w(\cdot)\geq 0} &\quad \sum_{k=1}^K \bbE\left[\int_{D}^{D+\Gamma}\texttt{mse}^{[k]}(t,\tilde{S}^{[k]})dt\right]-\beta \bbE\left[\Gamma\right]\notag\\
    \text{s.t.~} &\quad \bbE\left[w\left(\tilde{Y}\right)\right] \geq \frac{1}{1-\epsilon}\left(\frac{K}{f_{\max}}-\frac{K}{\mu}\right),
\end{align}
from which the optimal solution of problem \eqref{sumMSE} is given by $\beta^*$ that uniquely solves $p(\beta^*)=0$ \cite{dinkelbach-fractional-prog}.

Now for a fixed $\beta$, one can follow a Lagrangian approach to show that the optimal solution of problem (\ref{opt_aux}) satisfies
\setcounter{equation}{32}
\begin{align}
  \sum_{k=1}^K \frac{\sigma_k^2}{2\theta_k}&\left( 1-\bbE\left[e^{-2\theta_k Y} \right]e^{-2\theta_k(w^*(z)+z)}\right)\notag\\
  &\qquad\qquad\qquad\qquad= \beta+\zeta(1-\epsilon)+\frac{\eta(y)}{f_{\tilde{Y}}(z)}, 
\end{align}
with $\eta(y)$ and $\zeta$ being Lagrange multipliers corresponding to the dual problem, and $f_{\tilde{Y}}(\cdot)$ is the probability density function of the total service time in the epoch. We define the left hand side of the above as $G_{\boldsymbol{\theta},\boldsymbol{\sigma}}(w^*(z)+z)$ given in the theorem, which is an increasing function. Thus, one can uniquely solve for $w^*(z)$ in terms of the Lagrange multipliers. 

We then make use of the complementary slackness conditions and some involved mathematical manipulations to characterize the effect of the Lagrange multipliers on the optimal solution. Specifically, we define the function $H(\cdot)$ to denote the average waiting time and characterize it using the sampling constraint (when binding). The function $H(\cdot)$ depends on the distribution of $\tilde{Y}$, given by a convolution of a random number (that is geometrically distributed) of the $\exp(\mu)$ distribution. This gives rise to the incomplete Gamma function used in \eqref{H_fn} and \eqref{F_fn}.

Finally, everything is combined by solving $p(\beta^*)=0$.
\end{Proof}

Theorem~\ref{thm1} shows that the sensor only takes a new sample in the epoch only if the previous epoch's {\it total} service time $\sim\tilde{Y}$ (of all processes) surpasses a certain threshold. It is emphasized in \eqref{waiting-policy-opt} that such threshold depends on the system parameters. This is highlighted in the next section.

\section{Numerical Results}
\begin{figure}[t]
\center
\includegraphics[scale=.52]{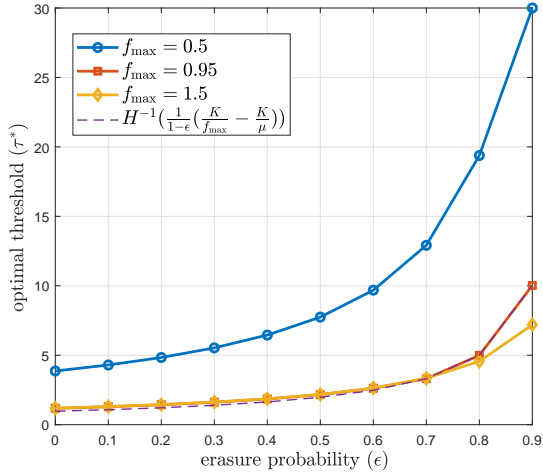}
\vspace{-.1in}
\caption{The optimal threshold ($\tau^*$) versus the erasure probability ($\epsilon$) for different sampling frequency constraints ($f_{\max}$).}
\label{fig:threshold_epsilon}
\vspace{-0.22in}
\end{figure}
In this section, we present our numerical results concerning Theorem~\ref{thm1}. In Fig.~\ref{fig:threshold_epsilon}, we study a 2-process system with $\boldsymbol{\theta}=[0.1 \quad 0.5]$, and $\boldsymbol{\sigma}=[1 \quad 2]$. The service rate $\mu=1$. We show the optimal threshold $\tau^*$ versus the erasure probability $\epsilon$ for $f_{\max}=0.5, \: 0.95, \:, 1.5$. The long-term average MMSE, $\beta^*$ increases as the erasure probability increases (omitted due to space limitations). Our results show that for all sampling frequency constraints, the optimal threshold increases as the erasure probability increases. This is due to the fact that the functions $G_{\boldsymbol{\theta},\boldsymbol{\sigma}}^{-1}(\cdot)$, and $H^{-1}(\cdot)$ are increasing functions in their argument, which is in turn is an increasing function in $\epsilon$. Nevertheless, we have three different cases. First, when $f_{\max}=0.5$, the sampling frequency constraint is binding even at $\epsilon=0$. Hence, the optimal threshold is given by $H^{-1}\left(\frac{1}{1-\epsilon}\left[\frac{K}{f_{\max}}-\frac{K}{\mu}\right]\right)=H^{-1}\left(\frac{2}{1-\epsilon}\right)$. Thus, the optimal threshold is higher than the remaining cases and much steeper. On the other hand, when $f_{\max}=1.5$, the sampling frequency constraint is inactive as $f_{\max}>\mu$, and the optimal threshold is given by $G^{-1}_{\boldsymbol{\theta},\boldsymbol{\sigma}}(\beta^*)$ for all $\epsilon$. Finally, for the case when $f_{\max}=0.95$, we observe an interesting behavior. When $\epsilon<\epsilon^*=0.7$, the threshold corresponding to $G^{-1}_{\boldsymbol{\theta},\boldsymbol{\sigma}}(\beta^*)$ is (slightly) higher than the threshold corresponding to  $H^{-1}\left(\frac{1}{1-\epsilon}\left[\frac{K}{f_{\max}}-\frac{K}{\mu}\right]\right)$ (which is shown as a dotted curve in Fig.~\ref{fig:threshold_epsilon}), while for $\epsilon>\epsilon^*=0.7$, the sampling frequency constraint becomes binding and therefore, the optimal threshold is characterized by $H^{-1}(\cdot)$ and becomes more steeper.
\begin{figure}[t]
\center
\includegraphics[scale=.52]{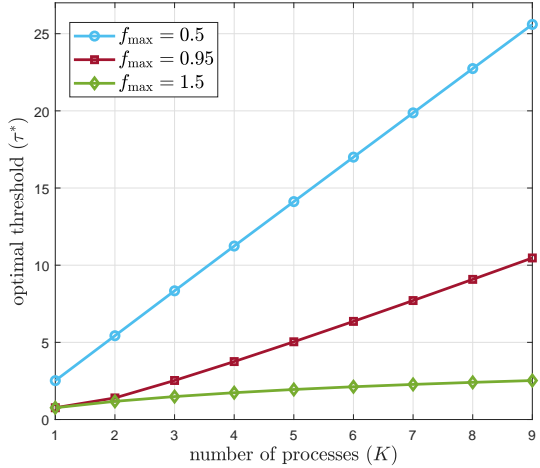}
\vspace{-.1in}
\caption{The optimal threshold ($\tau^*$) versus the number of processes ($K$) for different sampling frequency constraints ($f_{\max}$).}
\label{fig:threshold_users}
\vspace{-0.22in}
\end{figure}

In Fig.~\ref{fig:threshold_users}, we consider a symmetric system with $K$ processes, each having $\sigma_k^2=1$, and $\theta_k=0.5$ for all $k$ with service rate $\mu=1$. We study the optimal threshold variation with increasing the number of processes $K$. We observe that the long-term average sum MMSE increases with $K$ as expected (omitted due to space limitation).  Fig.~\ref{fig:threshold_users} shows that as $K$ increases, the optimal threshold increases as well. The slope of the curve depends on $f_{\max}$. When $f_{\max}=0.5$, the sampling frequency constraint is binding, and $\tau^*$ appears to linearly increase with $K$ with a steeper slope. When $f_{max}=1.5>\mu=1$, i.e., the problem becomes unconstrained, the optimal threshold is slowly increasing with $K$. For $f_{\max}=0.95$, the optimal threshold matches the unconstrained solution for $K=1,2$. Nevertheless, when $K>2$, the sampling frequency constraint becomes binding and the linear-like profile of the optimal threshold prevails. 

\begin{figure}[t]
\center
\includegraphics[scale=.52]{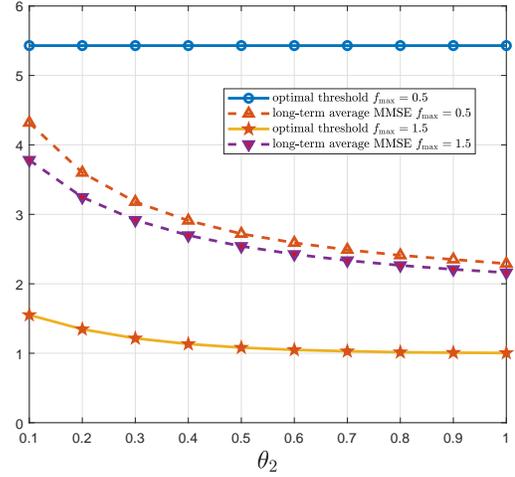}
\vspace{-.1in}
\caption{The optimal threshold ($\tau^*$) and the optimal long-term average MMSE versus $\theta_2$ for tracking two processes, where the first process has fixed parameters $\sigma_1^2=1$ and $\theta_1=1$ with different sampling frequency constraints ($f_{\max}$).}
\label{fig:threshold_theta}
\vspace{-0.22in}
\end{figure}

In Fig.~\ref{fig:threshold_theta}, we consider a 2-process system with $\sigma_1^2=2$, $\sigma_2^2=1$ and $\theta_1=0.5$. We vary $\theta_2 \in [0.1,1]$ and observe its effect on the optimal threshold and the MMSE for the same service rate $\mu=1$. We observe that when the sampling frequency constraint is binding, e.g., when $f_{\max}=0.5$, the optimal threshold is independent of $\theta_2$ as the argument of $H^{-1}(\cdot)$ is independent of $\theta_2$. The optimal threshold, however, is a monotonically decreasing function in $\theta_2$ for $f_{\max}=1.5$ as the process becomes faster and thus the system needs to wait less to track the variations in the process as long as the sampling constraint is inactive. In both cases, the long-term average MMSE is decreasing in $\theta_2$ since the sum of the processes' variances decreases.

\newpage
\bibliographystyle{unsrt}
\bibliography{reference}
\end{document}